\documentclass[useAMS,usenatbib,A4]{mn2e}
\usepackage{epsfig,amssymb,amsmath}
\usepackage{graphicx,amscd,amsmath,amssymb,verbatim}
\usepackage[TS1,OT1,T1]{fontenc}
\title[A Spectroscopic Study of S~Equ and KO~Aql]
{A Spectroscopic Study of the Algol-type Binaries S~Equulei and KO
Aquilae: Absolute Parameters and Mass Transfer\thanks{Based on
observations collected at the Catania Astrophysical Observatory
(Italy)}}

\author[Soydugan et al.]
{F. Soydugan$^{1}$ \thanks{e-mail: fsoydugan@comu.edu.tr}, A.
Frasca$^{2}$, E. Soydugan$^{1}$, S. Catalano$^{2}$, O.
Demircan$^{1}$, and C. {\.I}bano\v{g}lu$^{3}$ \\
$^{1}$ \c{C}anakkale Onsekiz Mart University, Faculty of Art and Science, Department of Physics, 17100, \c{C}anakkale, Turkey \\
$^{2}$ INAF -- Catania Astrophysical Observatory, via S. Sofia, 78, I-95123, Catania, Italy \\
$^{3}$ Ege University, Science Faculty, Astronomy and Space Science Dept., 35100 Bornova, \.{I}zmir, Turkey}

\begin{document}

\date{Accepted ...  Received ... ; in original form ... }

\pagerange{\pageref{firstpage}--\pageref{lastpage}} \pubyear{2007}

\maketitle

\label{firstpage}

\begin{abstract}

We present and analyze high-resolution optical spectra of two Algol
binaries, namely S~Equ and KO~Aql, obtained with the echélle
spectrograph at Catania Astrophysical Observatory. New accurate
radial velocities for the hotter primary components are obtained.
Thanks to the cross-correlation procedure, we were able to measure,
for the first time to our knowledge, radial velocities also for the
cool secondary components of S~Equ and KO~Aql. By combining the
parameters obtained from the solution of the radial velocity curves
with those obtained from the light curve analysis, reliable absolute
parameters of the systems have been derived. The rotational velocity
of the hotter components of S~Equ and KO~Aql has been measured and
it is found that the gainers of both systems rotate about 30\%
faster than synchronously. This is likely due to mass transfer
across the Lagrangian $L_1$ point from the cooler to the hotter
component. The lower luminosity of the mass-gaining components of
these systems compared to normal main-sequence stars of the same
mass can be also an effect of the mass transfer. The H$\alpha$
profiles were analyzed with the ``synthesis and subtraction''
technique and reveal clear evidence of mass transfer and accretion
structures. In both systems, especially before the primary eclipses
and afterwards, we clearly observed extra-absorption lines. From the
integrated absorption and the radial velocity variations of these
features, we found that the mass accretion is very dense around the
impact region of the hotter components. A double-peaked emission in
the spectra of S~Equ was seen outside the eclipses. One of these
peaks is likely originated in a region between the center of mass
and the cooler component, which is occupied by the flowing matter.
Furthermore, the H$\alpha$ difference spectra of S~Equ and KO~Aql
display also emission features, which should be arising from the
magnetic activity of the cooler components.

\end{abstract}

\begin{keywords}
stars: binaries: eclipsing - stars: mass transfer - stars: individual: S~Equ, KO~Aql
\end{keywords}

\section{Introduction}
Algol systems are close, interacting binaries, which consist of a
cool, giant or subgiant secondary star of F-K spectral type that
fills in its own Roche lobe and transfers mass and angular momentum
to a hot and more massive B-A type main-sequence primary star. They
are an important astrophysical laboratory for the study of different
phenomena such as mass transfer and accretion, magnetic activity in
the late-type companion, angular momentum and orbit evolution. A
crucial need for the study of all these processes is the knowledge
of the basic stellar and orbital parameters. In the last decades,
the number of Algol-type binaries with reliable fundamental
parameters has considerably increased due to the availability of
high-resolution spectroscopic data analyzed with new powerful
techniques (e.g. cross-correlation, Popper $\&$ Jeong, 1994) to
measure radial velocities, and to the accurate analysis of
multi-band (from visible to near-IR wavelength) light curves with
sophisticated spot models (e.g., L\'{a}zaro et al. 2004).

Observational evidences of mass transfer activity and accretion
structures in Algols are: (i) the variations of the light and color
curves, (ii) distortions of the radial velocity curves of the
primary components, (iii) the mass-gainer hotter components rotating
faster than synchronously, and (iv) extra emission and/or absorption
features in strong spectral lines such as H$\alpha$ and H$\beta$.
Some theoretical works have tackled the problem of Roche overflow
and mass transfer, predicting the path of the gas stream and the
accretion structures around the primary component (see, e.g.,
Prendergast \& Taam 1974, Lubow \& Shu 1975, Peters 1989).

In more recent years, different spectral behaviors (especially extra
H$\alpha$ emission and/or absorption features) have been observed
for short- and long-period Algol systems and have been attributed to
different effects of mass transfer due to its strength and to the
system geometry (Richards \& Albright 1999). In long-period systems
(P > 6 days), the gas stream misses the primary component and give
rise to a "permanent accretion disk", because the size of the mass
gainer is very small relative to the binary separation. The mass
transfer and accretion process is more complex in short-period
Algols (P $\leq$ 6 days). In this case, the stream particles mostly
impact the primary's photosphere due to the large size of the
primary relative to the separation of the components. The accretion
structure is called a "transient accretion disk" in Algols with
orbital periods in the range 4.5--6 days (Kaitchuck $\&$ Honeycutt
1982) and  an "accretion annulus" in the systems with P$_{orb}$ < 4.5
days (Richards et al., 1996). A wide spectroscopic survey of short-
and long-period Algol systems aimed at the study of mass transfer
has been performed by Richards $\&$ Albright (1999) and Vesper et
al. (2001). The two Algol systems studied by us in the present work
are both short-period ones.

S~Equulei (BD +04\degr  4584, HD 199454, HIP 103419, V$\simeq
8\fm$4) is a semidetached Algol-type eclipsing binary with an
orbital period of P = 3.4361 days. The  hotter and cooler components
have been classified as B9.5 V and F9 III-IV, respectively, by Plavec
(1966) and Plavec $\&$ Polidan (1976).

The first complete light curve was presented by Catalano \& Rodon\`o
(1968) who also obtained the light curve solution. They found the
system to be partially eclipsing and observed some peculiarities in
the light curve. In particular, they noticed a decrease in
luminosity, especially in $B$ light (about $0\fm04$), occurring near
0.9 phase. They interpreted this anomaly as due to the gaseous
stream. Piotrowski et al. (1974) have discussed the distortions of
the light curves of some semi-detached systems, including S~Equ with
the Catalano \& Rodon\`o data. They ascribed the brightening
observed in several systems just after the primary minimum to
emission from the gas stream that is observed along its axis, in the
direction of higher optical depth. The luminosity decrease observed
before the primary minimum is attributed to absorption by electron
scattering produced by the stream itself that is projected against
the disk of the hotter component at these phases. A more recent
light curve solution (Zola, 1992) suggests a third light comparable
to the contribution of the cooler component to the total flux of the
system.

Richards $\&$ Albright (1999), Vesper et al. (2001), and Richards (2001)
have studied S~Equ spectroscopically and found single-peaked H$\alpha$ emission
features with additional weak double-peaked emission at some epochs, indicating
mass transfer activity and a transient disk structure around the primary component.
Doppler tomography based on the H$\alpha$ differences
spectra displays extra-absorption and/or emission due to the gas
stream, accretion structures, and chromospheric activity associated with the cooler
component.

A detailed O--C analysis of the times of minima of S~Equ was recently
performed by Soydugan et al. (2003).
They found an increase in the orbital period of the system with a rate of
0.102 s\,yr$^{-1}$, requiring a mass transfer rate
$dM/dt \simeq 3.97 \times 10^{-8}$ M$_{\sun}$ yr$^{-1}$ from the secondary
to the primary component.

    KO~Aquilae (BD +10\degr 3655, HD 92177, V$\simeq 8\fm$4) is also a
classical Algol-type binary. Its variability was discovered by Hoffmeister
(1930) and its orbital period was determined as 2.864071 days by Plaut (1932).
Sahade (1945) estimated the spectral type of the primary component as
A0. Blanco $\&$ Cristaldi (1974) and Mader $\&$ Angione (1996)
analyzed the light curves of the system and determined the stellar
parameters. In the latter study, an orbital period increase with a rate of
0.375 s\,yr$^{-1}$ was also determined and explained as due to
the mass transfer process. Sahade (1945) published the radial velocity curve
and orbit solution of the primary component using plate spectra.
He was not able to detect any sign of the cool component in his spectra.
Vesper et al. (2001) obtained only one spectrum in the H$\alpha$ region, at the
orbital phase 0.611, and observed extra absorption and emission
attributed to accretion structures around the primary
component.

The basic orbital and physical parameters of S~Equ and KO~Aql, as
compiled from the literature, are reported in Table 1.

    In this work, we present the results of the analysis of new high-resolution ech\'elle
spectra of S~Equ and KO~Aql. The observations and data reduction are
described in \S 2. We report the radial velocity measurements and
the first complete orbit solutions for the components of S~Equ and
KO~Aql in \S 3. The projected rotational velocities of the hotter
components are discussed in \S 4. The results of the application of
synthetic spectra to reproduce the photospheric contribution of both
components, to refine the spectral classification, and to study the
H$\alpha$ extra-absorption and emission features are presented in \S
5 and \S 6. Finally, the results are discussed in \S 7 and
summarized in \S 8.

\begin{table*}
 \begin{center}
 \begin{minipage}{130mm}
  \caption{Basic parameters of S~Equ and KO~Aql}
  \begin{tabular}{@{}llllccccc@{}}
  \hline
   System & Spectral  & P$_{{orb}}$ & i & T$_{1}$   & T$_{2}$ &  r$_{1}$
     & r$_{2}$  & Ref. \\
     & Type& (days) & (deg) & (K) & (K) & (=R$_{1}$/a) & (=R$_{2}$/a) & \\

 \hline
 S~Equ  & B8-9.5V + F9III-IV & 3.436128 & 87.25 & 11200 & 5255 & 0.185  &  0.219 &  1, 2 \\
 KO~Aql & A0 + ...           & 2.863965 & 77.86 & 9900  & 4425 & 0.144  &  0.277 &  3, 4 \\
\hline
\end{tabular}
\end{minipage}

\begin{minipage}{125mm}
\begin{list}{}{}
\item $^{1}$Zola (1992), $^{2}$Kreiner (2004), $^{3}$Sahade (1945), $^{4}$ Mader $\&$ Angione (1996)
\end{list}

\end{minipage}
\end{center}
\end{table*}

\section[]{Observations and Data Reduction}

The spectroscopic observations have been performed at the Catania Astrophysical Observatory
with the echélle spectrograph (FRESCO), which is fiber-linked to the 91-cm telescope. The
data were obtained during two observing runs, with observing nights spread from 12 September to 19 November 2003
and from 12 April to 29 July 2004. The spectra
were obtained using the echélle configuration based on a 300 lines/mm echellette grating as
cross-disperser element and an echelle grating with 79 lines/mm. A CCD camera with a thinned
back-illuminated SITe chip of 1024$\times$1024 pixels (pixel size 24 $\mu$m) refrigerated to
$-130\degr$C by liquid nitrogen was used to acquire the data.
The spectra have a resolution R $\simeq$ 21\,000 and cover the wavelength range from 4300 to 6650 Å
separated into 19 orders. We have obtained 27 and 37 spectra for S~Equ and KO~Aql, respectively.
The integration times for the variables were in the range 40--55 minutes, and
the S/N ratio reached was typically between 50 and 80 at the continuum near the H$\alpha$, depending
on the weather conditions. More information about the observations can be found in Table 2.

In addition to S~Equ and KO~Aql, we also observed some reference
stars, whose spectral types are similar to those of the components of
the variables. We have chosen the non-active and slowly-rotating standard
stars 10 Tau (F9 IV-V) and $\alpha$ Boo (K1.5 III) to mimic the
cooler components of S~Equ and KO~Aql, respectively. The bright, slowly-rotating
stars, Vega (A0 V, V$_{r}= -13.8$ km\,s$^{-1}$), $\beta$ Vir (F9 V, V$_{r} = +4.6$ km\,s$^{-1}$), 
5 Ser (F8 III-IV, V$_{r} = +53.5$ km\,s$^{-1}$) and
$\alpha$ Boo (K1.5 III, V$_{r} = -5.2$ km\,s$^{-1}$) have been chosen as templates for
the radial velocity measurements of the hotter and cooler components and have been
observed several times during each run.

The data reduction was performed by using the ECHELLE task of
\texttt{IRAF\footnote{\texttt{IRAF} is distributed by the National
Optical Astronomy, which is operated by the Association of
University for Research in Astronomy, inc. (AURA) under cooperative
agreement with the National Science Foundation.}} package following
the standard steps: background subtraction, division by a flat field
spectrum given by a halogen lamp, wavelength calibration using the
emission lines of a Thorium-Argon lamp, and normalization through a
polynomial fit to the continuum.

\begin{table}
 \begin{center}
 \begin{minipage}{90mm}
  \caption{Information about the spectroscopic observations}
  \begin{tabular}{@{}lcccc@{}}
  \hline\hline
              & Observing  & Number    & Exposure  & Number \\
  System     & Runs$^{*}$ & of nights & time (s)  & of spectra  \\
 \hline
 S~Equ       & 1          & 8         & 2400--3300 & 11 \\
             & 2          & 15        & 2400--3300 & 16 \\
 KO~Aql      & 1          & 10        & 2400--3300 & 11 \\
             & 2          & 20        & 2400--3300 & 26 \\
\hline
Standard Stars    &            &           &           &    \\
\hline
 Vega (A0V)    & 1          & 8         & 15--30     & 8  \\
               & 2          & 19        & 15--30     & 20 \\
 10 Tau (F9IV-V) & 1          & 2         & 600       & 2  \\
 $\beta$ Vir (F9V) & 2          & 10        & 300--600   & 10 \\
 5 Ser   (F8III-IV)    & 2          & 2         & 720       & 2  \\
 $\alpha$ Boo (K1.5III)& 2          & 30        & 20--30     & 35  \\

\hline\hline
\end{tabular}
\end{minipage}

\begin{minipage}{70mm}
\begin{list}{}{}
\item $^{*}$Run 1: 12 September -- 19 November 2003\\
            Run 2: 12 April -- 29 July 2004

\end{list}
\end{minipage}
\end{center}
\end{table}

The red part of the spectra is affected by many water vapor and
O$_{2}$ telluric lines. We have removed these lines at the H$\alpha$
wavelengths using the spectra of Altair (A7 V, $v\sin i = 245$ km\,s$^{-1}$)
and $\alpha$ Leo (B7 V, $v\sin i =  353$ km\,s$^{-1}$)
acquired during the observing runs. These spectra have been
normalized, also inside the very broad H$\alpha$ profile, to provide
valuable templates for the water vapor lines. Applying an
interactive method that is described by Frasca et al. (2000), the
telluric lines were cleaned from the spectra of S~Equ and KO~Aql.

\section[]{Radial Velocities and Orbit Solutions}
The cross-correlation technique (CCT) has been widely and successfully used,
after the advent of electronic detectors, for the measurements of radial
velocity (RV) reaching precisions as small as a few tens of m\,s$^{-1}$
in particular cases, leading to the detection of giant planets around
solar-type stars (e.g., Mayor \& Queloz, 1995).

By means of CCT, RVs of many binary systems were obtained with
better precision and, in several single-lined binaries, the signal
of the previously unseen companion has been detected. In the case of
Algol-type binaries, the RV measurement of the cooler components is
normally hampered by the strong luminosity contrast between the
primary and secondary stars. The CCT has been proven to be very
powerful for detecting the RVs of the secondary components of Algol
systems (e.g., Holmgren 1988, Hill 1993, Khalesseh \& Hill 1992,
Maxted \& Hilditch 1995).

In the present work, RVs of both components of S~Equ and KO~Aql were
derived by means of the CCT using the \texttt{IRAF} task FXCOR (e.g.
Tonry \& Davis 1979, Popper \& Jeong 1994). The wavelength ranges
were selected to exclude Balmer and Na\,{\sc i} D$_2$ lines, which strongly broaden
the cross-correlation function (CCF) with their wings and can also be
contaminated by the mass transfer and chromospheric activity. The
spectral regions heavily affected by telluric absorption lines (e.g.
the O$_{2}$ band $\lambda$ 6276-6315) were not used. The
RVs of the components of S~Equ and KO~Aql, listed in Tables 3 and 4
together with their standard errors, are the weighted averages of
the values obtained from the cross-correlation of each order of the
target spectra with the corresponding order of the standard star
spectrum. The weight W$_{i}$ = 1/$\sigma_{i}^{2}$ has been given to
each measurement. The standard errors of the weighted means have been calculated
on the basis of the errors ($\sigma_{i}$) in the RV values for each
order according to the usual formula (e.g., Topping, 1972). The
$\sigma_{i}$ values are computed by FXCOR according to the fitted
peak height as described by Tonry \& Davis (1979).

Spectra of the RV standard stars $\alpha$ Boo (K1.5 III) and $\beta$ Vir (F9 V)
have been used as templates for deriving the radial velocity of the secondaries of
KO~Aql and S~Equ, respectively. An average high-S/N spectrum of Vega was instead used
as template for the measurement of the RV of the primaries.
As pointed out by Griffin (1999), it is not easy to achieve very accurate radial
velocity measurements (accuracy better than 100 m\,s$^{-1}$) for early type stars, due to the
small number of suitable lines spread in a wide wavelength range and to other factors 
affecting the early-type line spectra. She found that real stellar spectra can be used as
RV templates, because they are free from uncertainties in the tabulated wavelengths and
have line shapes and intensities that are actually encountered. From cross-correlation of
several high-resolution spectra of Vega with an average template built up with Sirius spectra,
she found RV errors of about 250--350 m\,s$^{-1}$. These errors, possibly systematic, could be 
related to the known low-amplitude $\delta$\,Scuti-type variability of Vega (e.g., Samus et al. 2004)
but are much smaller than the standard errors on the individual RVs of the primary components
of S~Equ and KO~Aql (typically from 1 to 7 km\,s$^{-1}$), and so they can be neglected.

\subsection[]{S~Equ}
The RV curve of the primary component of S~Equ has been obtained by
Plavec (1966) based on plate spectra, which were taken by Petrie and
Thackeray at Dominion Astrophysical Observatory and Radcliffe
Observatory. The orbital solution obtained by them and,
subsequently, by Lucy $\&$ Sweeney (1971) on the same very scattered
data is characterized by a relatively large eccentricity
($e\,\approx 0.15$). This rather high value is not supported by 
the light curves with equally spaced minima of the same duration observed
by Catalano \& Rodon\`o (1968) and Zola (1992).
Our RV data definitely exclude an eccentricity larger than about 0.02, and so  
a circular orbit was assumed in the following analysis.
Up to now, there has been no published RV curve of the secondary component, to
our knowledge.

\begin{table}
 \begin{center}
  \caption{Radial velocity measurements of S~Equ}
  \begin{tabular}{@{}cccr@{}}
  \hline
    HJD     & Orbital   & V$_{h}$          & V$_{c}$~~~~~       \\
  2\,450\,000+ & Phase     & (km s$^{-1}$)    & (km s$^{-1}$)  \\
  \hline

2895.4598   & 0.1907    & $-81.7 \pm$ 2.6   & $122.4 \pm$15.5   \\
2896.4473   & 0.4781    & $-55.8 \pm$ 3.0   &   --~~~~~~         \\
2910.4581   & 0.5556    & $-45.5 \pm$ 6.9   &   --~~~~~~         \\
2922.3630   & 0.0202    & $-79.8 \pm$ 5.6   &$ -32.3 \pm$ 3.8    \\
2922.4364   & 0.0416    & $-66.1 \pm$ 6.5   &$ -12.6 \pm$ 4.2    \\
2923.4058   & 0.3237    & $-72.8 \pm$ 9.1   &$ 121.0 \pm$ 7.1     \\
2924.4177   & 0.6182    & $-37.3 \pm$ 4.9   &$-198.9 \pm$ 4.7 \\
2933.3604   & 0.2207    & $-79.3 \pm$ 7.1   &$ 131.6 \pm$ 4.7     \\
2934.2877   & 0.4906    & $-50.2 \pm$ 5.1   &$ -39.1 \pm$15.0    \\
2934.3345   & 0.5042    & $-49.9 \pm$ 3.8   &$ -44.6 \pm$ 10.8     \\
2934.3619   & 0.5122    & $-49.3 \pm$ 4.6   &$-102.8 \pm$ 8.1  \\
3155.5306   & 0.8779    & $-38.2 \pm$ 1.3   &$-184.2 \pm$ 2.9  \\
3157.5783   & 0.4738    & $-56.1 \pm$ 2.0   &  --~~~~~~      \\
3158.5586   & 0.7591    & $-28.5 \pm$ 1.9   & $-247.2 \pm$ 3.4  \\
3160.5341   & 0.3340    & $-74.5 \pm$ 2.6   & $ 116.7 \pm$ 3.5     \\
3166.5869   & 0.0955    & $-68.5 \pm$ 1.9   & $  55.7 \pm$ 3.1    \\
3167.5542   & 0.3770    & $-70.9 \pm$ 1.4   & $  72.0 \pm$ 2.6    \\
3168.5489   & 0.6665    & $-32.2 \pm$ 2.3   & $-227.4 \pm$ 3.7  \\
3170.5519   & 0.2494    & $-79.3 \pm$ 1.7   & $ 135.7 \pm$ 3.7     \\
3171.5905   & 0.5517    & $-49.5 \pm$ 1.9   & $-123.7 \pm$ 8.0  \\
3185.5247   & 0.6069    & $-37.2 \pm$ 2.5   & $-185.1 \pm$ 4.5  \\
3186.4374   & 0.8725    & $-36.7 \pm$ 1.7   & $-184.5 \pm$ 3.8  \\
3202.5972   & 0.5754    & $-43.2 \pm$ 1.5   & $-152.0 \pm$ 5.2  \\
3203.4302   & 0.8178    & $-30.9 \pm$ 1.6   & $-226.0 \pm$ 4.0  \\
3213.4549   & 0.7353    & $-29.6 \pm$ 1.9   & $-243.5 \pm$ 5.0  \\
3214.4695   & 0.0290    & $-65.0 \pm$ 1.6   & $ -21.6 \pm$ 2.9     \\
3214.5830   & 0.0636    & $-63.7 \pm$ 1.7   & $  22.4 \pm$ 2.2    \\
\hline
\end{tabular}
\end{center}
\end{table}

In Figure~1, the measured RVs and the associated error bars of the
components of S~Equ are plotted as a function of the orbital phase.
Filled and open circles represent the primary and secondary
components, respectively. The orbital solution (dashed line for the
cooler and solid line for the hotter component) was determined,
assuming circular orbits, by a least-squares fit which reproduces
very well all the observed RVs.
 We report in Table 5 the orbital parameters of the system derived from
the solution of our radial velocity curves and compare them with with 
the values derived by Plavec.

\begin{figure}
\label{fig:RV_SEqu}
\begin{center}
\includegraphics[width=90mm,height=70mm]{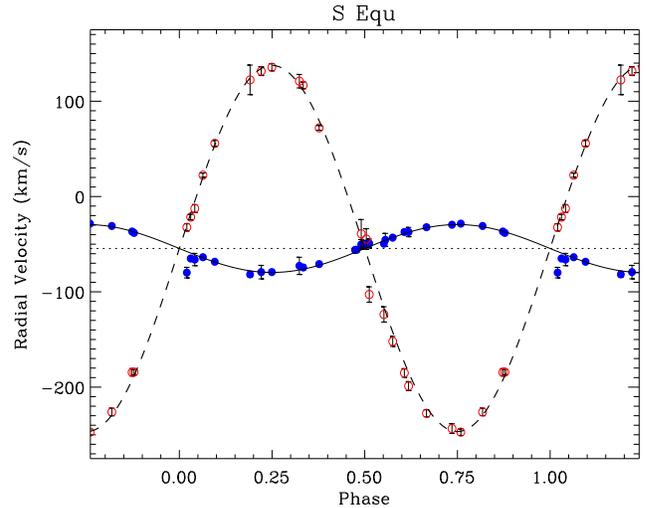}
\caption{Radial velocity curves of a double-lined orbit for S~Equ
plotted as a function of orbital phase. Filled and open circles
represent the RVs of primary and secondary components, respectively.
Solid line represents the orbital solution for the more massive
component, while the dashed line for the less massive component.}
\end{center}
\end{figure}

In the orbital solution of Plavec (1966) given in Table 5, the
systemic velocity, $\gamma$, and the semi--amplitude of the RV curve, $K_1$,
were found to be $-47.95 \pm 0.6$ km\,s$^{-1}$ and $23.4 \pm 0.8$ km\,s$^{-1}$,
respectively. 

\subsection[]{KO~Aql}

Sahade (1945) obtained the first RV measurements and orbital
solution for the primary component of KO~Aql based on low-resolution
spectra. He was not able to detect the secondary component due to
its very low contribution to the spectrum of the system observed at
blue wavelengths. Then, Lucy $\&$ Sweeney (1971) performed a new
orbital solution for the primary component with results similar to
those of Sahade (1945). We have obtained high-resolution spectra
that allowed us to determine the orbital parameters of the cooler
component for the first time and those of the hotter one with a
better precision than in previous studies. The RVs were measured
from the observed spectra using the cross-correlation technique as
described above. Two template spectra, Vega for primary component
and $\alpha$ Boo for secondary component, have been used for the
CCT. The final RVs for KO~Aql are listed in Table~4. The larger
errors in the velocities of the secondary (up to 3--4 times the
primary's ones) are due to the very low contribution of the
secondary star to the observed spectrum and to its high rotational
velocity that broaden and weaken its spectral lines. The fit of an
eccentric solution to the observed RVs of both components provided
very low, not-significant values of eccentricity ($e \approx 0.01$). 
Thus, a circular
orbit was assumed for the fit of the observed RV curves, obtaining
the elements reported in Table 5, where the ephemeris was adopted
from Kreiner (2004).

A plot of the observed RVs together with the fitted RV curves is
shown in Fig.~2. The orbital solutions determined by Sahade (1945)
for only the primary component and that obtained in this work (see Table
5 for comparison) agree very well with each other.

\begin{table}
 \begin{center}
  \caption{Radial velocity measurements of KO~Aql}
  \begin{tabular}{@{}ccrr@{}}
  \hline
    HJD     & Orbital   & V$_{h}$~~~~          & V$_{c}$~~~~       \\
  2\,450\,000+ & Phase     & (km s$^{-1}$)    & (km s$^{-1}$)  \\
  \hline

2895.4293   & 0.4696    & $ -7.1 \pm$ 4.2   & --~~~~~   \\
2896.3330   & 0.7851    & $ 42.7 \pm$ 3.4   & --~~~~~   \\
2910.3800   & 0.6897    & $ 39.2 \pm$ 2.1   & $-164.4 \pm$ 7.2  \\
2922.2714   & 0.8416    & $ 33.5 \pm$ 0.3   & $-142.3 \pm$ 5.8  \\
2923.2988   & 0.2003    & $-32.3 \pm$ 1.2   & $165.3  \pm$ 3.6   \\
2924.3282   & 0.5598    & $ 12.8 \pm$ 1.1   & $-62.3  \pm$ 7.8  \\
2925.3135   & 0.9038    & $ 22.4 \pm$ 1.5   & $-85.1  \pm$ 4.7  \\
2933.3237   & 0.7006    & $ 39.5 \pm$ 1.5   & $-165.4 \pm$ 6.1  \\
2934.2506   & 0.0242    & $-15.2 \pm$ 1.5   & $21.0   \pm$ 3.8    \\
2959.2777   & 0.7625    & $ 41.9 \pm$ 3.8   & $-173.8 \pm$ 9.1  \\
3125.5480   & 0.8164    & $ 37.6 \pm$ 1.5   & $-155.4 \pm$ 3.6  \\
3126.4985   & 0.1483    & $-29.0 \pm$ 1.3   & $146.6  \pm$ 2.7   \\
3127.5205   & 0.5051    & $  0.9 \pm$ 1.1   &  --~~~~~		 \\
3137.5032   & 0.9906    & $  8.0 \pm$ 2.2   & $-7.5   \pm$ 2.1   \\
3137.5437   & 0.0047    & $  1.2 \pm$ 2.1   & $-0.9   \pm$ 2.8   \\
3138.5306   & 0.3493    & $ -26.5\pm$ 3.4   & $145.3  \pm$ 1.7   \\
3139.5625   & 0.7096    & $ 36.5 \pm$ 1.4   & $-170.9 \pm$ 4.3  \\
3140.4825   & 0.0308    & $-12.4 \pm$ 1.4   & $31.9   \pm$ 3.9    \\
3140.5669   & 0.0603    & $-11.0 \pm$ 1.4   & $66.8   \pm$ 5.7    \\
3143.4979   & 0.0837    & $-17.0 \pm$ 1.5   & $86.8   \pm$ 4.0    \\
3143.5831   & 0.1134    & $-23.7 \pm$ 1.5   & $109.7  \pm$ 2.2   \\
3152.4905   & 0.2235    & $-36.5 \pm$ 1.1   & $173.4  \pm$ 5.4   \\
3152.5880   & 0.2575    & $-34.6 \pm$ 1.1   & $181.0  \pm$ 3.1   \\
3155.5687   & 0.2892    & $-34.7 \pm$ 0.9   & $166.5  \pm$ 4.9   \\
3156.5837   & 0.6526    & $ 32.9 \pm$ 1.0   & $-144.5 \pm$ 3.0  \\
3158.5267   & 0.3310    & $-31.7 \pm$ 0.9   & $154.4  \pm$ 5.4   \\
3161.5530   & 0.3877    & $-20.6 \pm$ 1.5   & $123.8  \pm$ 3.3   \\
3166.5406   & 0.1291    & $-25.0 \pm$ 1.2   & $124.2  \pm$ 6.0   \\
3167.4317   & 0.4403    & $ -9.7 \pm$ 1.9   & $78.0   \pm$ 6.2  \\
3167.5478   & 0.4808    & $ -0.1 \pm$ 0.7   & $26.9   \pm$ 2.7  \\
3168.4281   & 0.7882    & $ 39.9 \pm$ 1.3   & $-165.1 \pm$ 3.3  \\
3168.5833   & 0.8423    & $ 31.7 \pm$ 0.9   & $-141.0 \pm$ 3.1  \\
3170.5860   & 0.5416    & $ 10.2 \pm$ 0.7   & $-51.2  \pm$ 3.7  \\
3171.5641   & 0.8831    & $ 27.8 \pm$ 0.9   & $-111.7 \pm$ 2.6  \\
3183.4394   & 0.0294    & $-12.4 \pm$ 0.9   & $26.9   \pm$ 2.2    \\
3216.5501   & 0.5901    & $ 22.4 \pm$ 1.0   & $-98.4  \pm$ 2.7  \\
\hline
\end{tabular}
\end{center}
\end{table}

\begin{figure}
\begin{center}
\includegraphics[width=90mm,height=70mm]{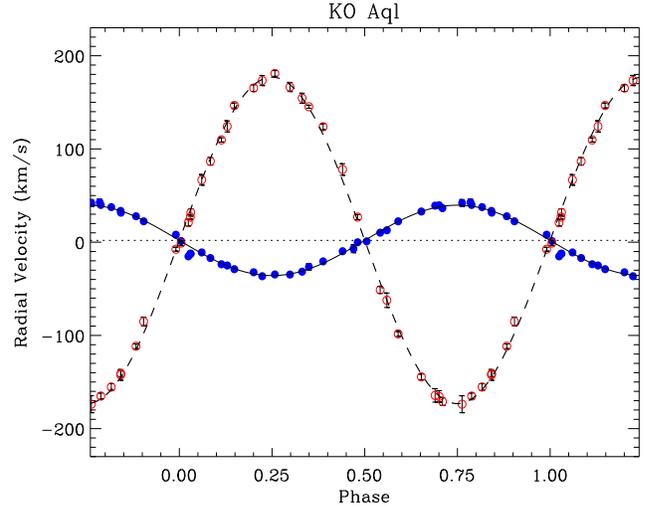}
\caption{Radial velocity curves of the components of KO~Aql plotted
as a function of orbital phase. The symbols and
lines have the same meaning as in Fig.~1.}  
\end{center}
\label{fig:RV_KOAql}
\end{figure}

\begin{table*}
  \caption{Spectroscopic orbital elements of S~Equ and KO~Aql.}
  \label{tab:orbital}
  \begin{center}
    \leavevmode
    \footnotesize
    \begin{tabular}[h]{lllllllll}
      \hline\hline \\[-5pt]
                     & & \multicolumn{3}{c} {S~Equ}    & &  \multicolumn{3}{c} {KO~Aql}    \\[+5pt]
       Parameter    & & Plavec (1966) & & This study    & & Sahade (1945) &  & This study        \\[+5pt]
      \cline{1-1}\cline{3-5}\cline{7-9}\\
      T$_{0}$ (HJD)   & &  2437568.345 & & 2452503.086\footnotemark[1] & & 2426585.442 & & 
2452501.707\footnotemark[1]  \\
      P$_{orb}$ (day) & & 3.4360726  & & 3.436128\footnotemark[1]    & & 2.863844    & & 2.864068\footnotemark[1]  \\
      $e$    & & 0.14  & & $0.01\pm0.01$\footnotemark[2]  & & 0.02   & & $0.01\pm0.01$\footnotemark[2]  \\
      $\gamma$ (km s$^{-1}$) & & $-47.95$& & $-54.5 \pm$ 0.5            & & $-2.7$        & & 1.9 $\pm$ 0.1       \\
      K$_{1}$  (km s$^{-1}$) & & 23.4& & 25.1 $\pm$ 0.6              & & 37.8        & & 37.8 $\pm$ 0.2        \\
      K$_{2}$  (km s$^{-1}$) & &     & & 192.1 $\pm$ 1.2             & &             & & 175.1 $\pm$ 0.9   \\
    a$_{1}$ sin i (10$^{6}$\,km) &  & 1.06  & & 1.19 $\pm$ 0.03      & &  1.5        & & 1.49 $\pm$ 0.01 \\
    a$_{2}$ sin i (10$^{6}$\,km) &  &       & & 9.08 $\pm$ 0.06      & &             & & 6.90 $\pm$ 0.03  \\
      M$_{1}$ sin$^{3}$\,i (M${_{\sun}}$) & & & & 3.23 $\pm$ 0.06    & &             & & 2.36 $\pm$ 0.03       \\
      M$_{2}$ sin$^{3}$\,i (M${_{\sun}}$) & & & & 0.42 $\pm$ 0.01    & &             & & 0.51 $\pm$ 0.01        \\
      q (=M$_{2}$/M$_{1}$)  &             & & &    0.131 $\pm$ 0.003 & &             & & 0.216 $\pm$ 0.002      \\
         \hline\hline \\
     \end{tabular}\\[-7pt]
      \end{center}
      \footnotemark[1]{The ephemeris was adopted from Kreiner (2004). }\\
      \footnotemark[2]{$e=0$ was assumed. }\\
      \end{table*}

\section[]{Rotational Velocities}

The synchronization time is shorter than the duration of the semi-detached
phase for Algol-type binaries by two or three orders of magnitude. In this
case, one should expect that the number of Algol systems with a non-synchronous
rotation must be much smaller than that of synchronous systems (Glazunova 1999).
However, the measured rotation rates of the primary components of Algol systems
indicate that most of them are rotating faster than in the case of synchronization
(Van Hamme \& Wilson 1990). The mass transfer process from the secondary
components to the primary ones can cause a strong increases in the
rotational velocity of the mass-gaining components because the matter flowing
through the L$_1$ Lagrangian point also transfers angular momentum (see, e.g.,
Giuricin et al. 1984, De Greve at al. 2006, Van Resbergen et al. 2006).

Van Resbergen et al. (2006) have stated that a ``conservative''
evolution (without loss of mass and angular momentum) fails to
explain the observed mass-ratios ($q=0.4$--1) while models with some
mass and angular momentum loss from these systems can better explain
the observed mass-ratios. They have proposed a scenario in which the
gainer is spun up by the incoming mass from the donor. The gainer
spins up - sometimes up to the critical velocity - and spins also
down due to tidal interaction. Even when the gainer has the critical
velocity this kinetic energy is not sufficient to drive mass out of
the system, as a consequence of the virial theorem. The extra-energy
needed is given by the accretion luminosity which is created in a
hot spot of the gainer's surface. As a consequence, they found that,
in massive Algol systems (with a donor starting with an initial mass
$\geq\,5\,M_{\sun}$), the combined action of rapid rotation and
accretion luminosity can overrun the binding energy of material on
the gainer's surface. The less-massive Algol systems (like, e.g.,
RZ~Cas) are instead likely obtained through an evolution without
loss of mass but with a considerable loss of angular momentum (e.g.
due to magnetic braking).

Therefore, the $v\sin i$ of the accreting components represents 
observational evidence for mass transfer and accretion phenomena.
Moreover, the values of $v\sin i$ are needed to produce correct
synthetic spectra.

We measured the projected rotational velocity ($v\sin i$) of
the ``mass gainers'' (i.e. hotter components) of S~Equ and KO~Aql by using the
CCF analysis. For this purpose, we obtained a calibration relation between
the full width at half maximum (FWHM) of the CCF peak and the $v\sin i$ with
artificially broadened spectra of slowly-rotating B9-A0 stars (Vega, 21~Peg,
and HR~6096) that were acquired in the same observing runs and with the same
instrumental setup of S~Equ and KO~Aql observations. Then,
the rotation velocities of the hotter components of the systems were
determined by converting the FWHM of its CCF peaks into $v\sin i$
through the aforementioned calibration. We found, as weighted averages of all
measurements, $v\sin i= 52.4\,\pm$\,4.4 km\,s$^{-1}$ and 41.0\,$\pm$\,2.0 km\,s$^{-1}$
for S~Equ and KO~Aql, respectively. The rotational velocity for S~Equ is in
agreement, within the errors, with the value of $v\sin i= 47.4\,\pm$\,1.4 km\,s$^{-1}$ 
determined by Mukherjee et al. (1996).

We have also calculated the rotational velocities of both components
of the systems in the hypothesis of synchronous rotation, finding
$v_{\rm sync}\sin i= 40$km\,s$^{-1}$ and 30 km\,s$^{-1}$ for the
hotter components of S~Equ and KO~Aql, respectively (see, Table 9).
The measured $v\sin i$ values for the gainers indicate that they are
spinning faster than the synchronous rate.

\section[]{Synthetic Spectra and H$\alpha$ Difference Profiles}

The emission and absorption features produced by accretion around
the hotter component in short-period Algol systems are generally
weak relative to the continuum flux of the spectra. In order to
emphasize the effect of mass transfer (and/or other physical
processes such as magnetic activity in cool components) in
H$\alpha$, it is necessary to remove the strong photospheric
contribution of both components from the observed spectra. Hence, we
have produced synthetic spectra for the hotter components built up
with Kurucz (1979) model atmospheres. Standard stars' spectra have
been used for the cooler components. For this purpose, the low-activity
stars 10 Tau (F9 IV-V) and $\alpha$ Boo (K1.5 III) for S~Equ and
KO~Aql, respectively, were observed in the same observing runs and
with the same instrumental setup.

Before forming the difference profiles (observed - synthetic) for the
H$\alpha$ line, we have checked the physical parameters (T$_{\rm
eff}$ and $\log g$) of the hotter components reported in the
literature. This task was accomplished by fitting Kurucz (1979)
model atmosphere spectra to the observed ones in spectral regions
where hot components overwhelm the secondary ones and mass transfer
effects are not visible. We have chosen the blue region around 4500
\AA ~and H$\gamma$. We have also taken into account the
small contribution (from 3 to 7 \%) of the cooler components at these
wavelengths by adding observed spectra of cool standard stars
properly weighted, broadened and Doppler-shifted. The temperature of
the primary component of S~Equ found using this method, $T_1=11,500$\,K, is only slightly
higher (300\,K) than that given by Zola (1992). For KO~Aql, we find the
same temperature ($T_1=9900$\,K) found by Mader \& Angione (1996).

The stellar parameters resulting from these fits are listed in Table
6, which contains also other parameters used to create the synthetic
spectra for S~Equ and KO~Aql. Some examples of observed and
synthetic spectra are displayed in Fig.~3. Then, we have formed the
composite synthetic spectra in the H$\alpha$ region to be compared
with the observed ones and to produce difference profiles. Effective
temperatures and absolute radii (in Table 9) were used to calculate
the contribution of primary and secondary components, whose spectra
were also rotationally broadened and wavelength Doppler-shifted,
before being summed. The synthetic profiles were then subtracted
from the observed ones, and the resulting difference profiles were
examined.

\begin{table*}
 \begin{center}

  \caption{The model parameters and standard stars for S~Equ and KO~Aql to create synthetic spectra.}
  \begin{tabular}{@{}lccccccc@{}}
  \hline\hline
       & & \multicolumn{3}{c} {Hot Component}   & & \multicolumn{2}{c} {Cool Component} \\
System & & T   & log g & $v\sin i$     & & Standard & $v\sin i$ \\
       & & (K) &       & (km\,s$^{-1}$)& &          & (km\,s$^{-1}$)\\
\cline{1-1}\cline{3-5}\cline{7-8}\\
 S~Equ  & &11500 & 4.1 & 52                  & &10 Tau (F9 IV) &        48 \\
 KO~Aql & & 9900  & 4.3 & 41                  & &$\alpha$ Boo (K1.5 III) & 58 \\
\hline\hline\\
\end{tabular}
\end{center}
\end{table*}

\begin{figure}
\begin{center}
\includegraphics[width=70mm,height=45mm]{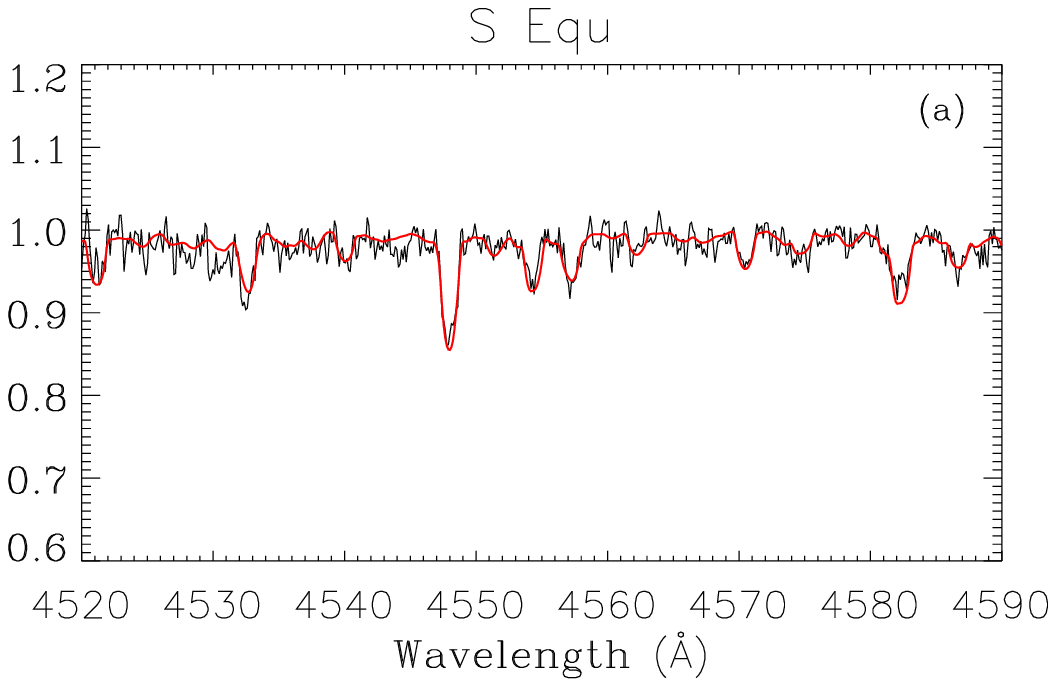}
\includegraphics[width=70mm,height=45mm]{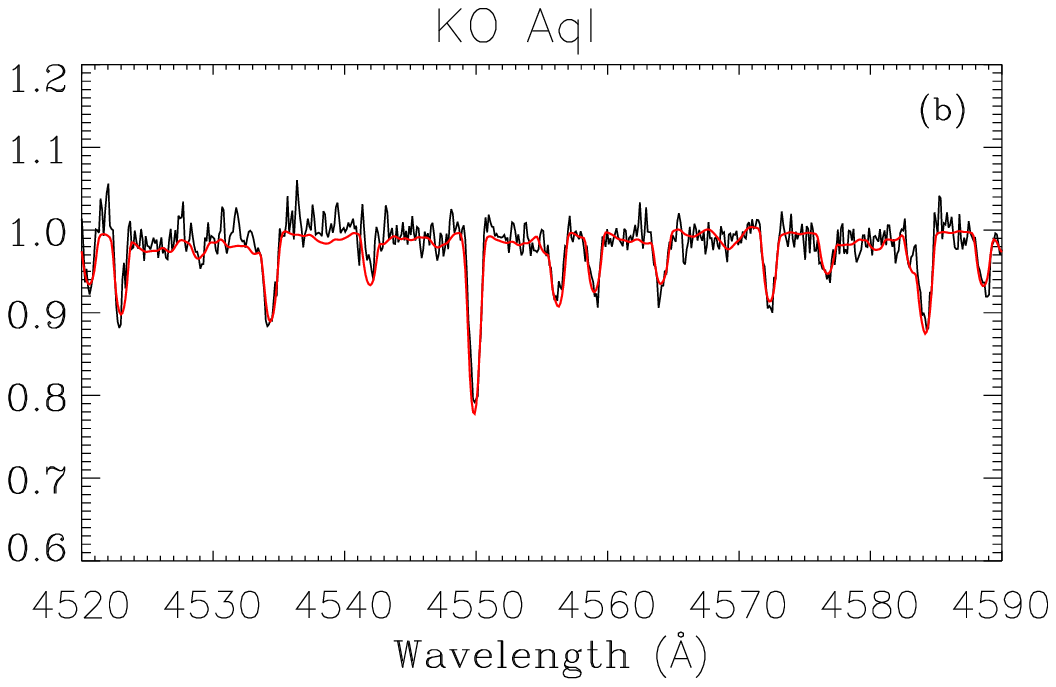}
\caption{Comparison between the observed (thin line) and the synthetic
spectra (thick line), normalized to a continuum of unity in the blue
spectral region that gives the better match for S~Equ (a) and KO~Aql
(b).}
\end{center}
\label{fig:spec_fit}
\end{figure}

\section[]{H$\alpha$ Extra-Emission and Absorption}

The H$\alpha$ residual profiles of Algol systems generally display
extra absorption and/or emission features commonly interpreted as
the result of Roche-lobe overflow, accretion structures around the
hotter components, and/or due to the chromospheric activity of the
cooler components. In short-period systems the contribution of the
active chromosphere of the rapidly-rotating cooler components can
produce significant effects on the H$\alpha$ line profile (Richards
$\&$ Albright 1999). The two Algol systems with relatively short
periods and with late-type secondary components that we are 
studying present both effects, namely mass transfer and
chromospheric activity.

\subsection[]{S~Equ}

The H$\alpha$ difference profiles of S~Equ display extra absorption
and emission that are variable with orbital phase. These spectral
features originate from accretion structure around the primary,
flowing gas in the region between the two stars and by the impact of
the gas stream on the primary component (Richards $\&$ Albright
1999, Vesper et al. 2001, Richards 2004). H$\alpha$ emission closely
associated with the late-type secondary component and attributed to
chromospheric activity has been also detected by Richards (2001) in
the H$\alpha$ tomogram of S~Equ.

In this work, we have analyzed H$\alpha$ difference profiles by
selecting the spectra with higher S/N ratio. Examples of the
observed, synthetic and difference H$\alpha$ profiles of S~Equ are
shown in Fig.~4. Several features could be readily noted. Some extra
emission is present at nearly all phases except near primary
eclipse. These emission components with single or double peaks are
more clearly visible around the quadratures and the stronger peaks,
which are red shifted, can be seen near phase 0.25. Although the
shape of the features at phase $\sim$\,0.75 is similar, the strength
is slightly weaker. The extra-absorption is more prominent just
before and after primary eclipse and it is still present at other
phases though with a decreasing strength.

\begin{figure}
\begin{center}
\includegraphics[width=85mm,height=140mm]{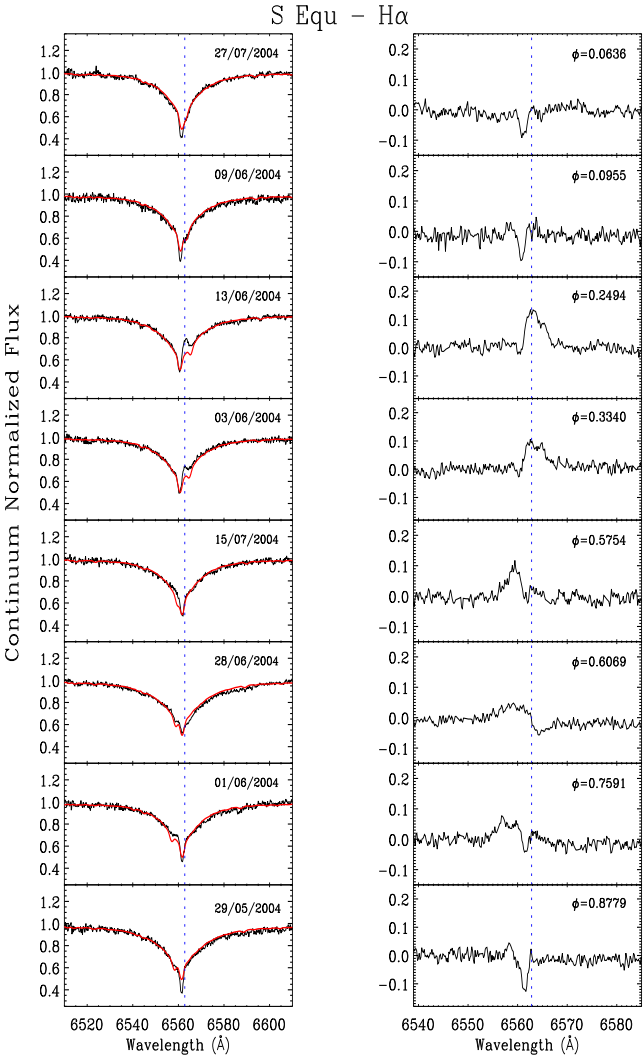}
\caption{\emph{Left.} Sample of observed (thin line) and the synthetic
spectra (thick line) for S~Equ at different orbital phases. The date of
observation is marked in each panel.
\emph{Right.} The difference spectra showing H$\alpha$ extra-absorption and emission features.
The orbital phase is reported in the upper right corner of each panel.
The vertical dashed lines in the lefts and right boxes  represent the laboratory wavelength ($\lambda_{0} = 6562.82\AA$) of H$\alpha$ line. }
\end{center}
\label{fig:Halpha_SEqu}
\end{figure}

Considering all the residual profiles obtained at different orbital
phases, a general structure, consisting of a weak absorption (A) and
broad, single- or double-peak emission (B and C), is met (see
Fig.~5).

With the aim of determining the sources of such features, we
measured the equivalent widths (EWs) and radial velocities (RVs) for
the absorption and emission components in the difference profiles by
means of multiple Gaussian fits. These values are reported in Table
7. Moreover, in some spectra we have detected other features which
seem to be unrelated to the ones already identified as A, B, and
C, on the basis of their RV. These unidentified features are
referred to simply as ``absorption'' or ``emission'' in Table~7.

The EW and RV values for the A, B, and C features are plotted versus
orbital phase in Fig.~6 with asterisks, plus signs, and triangles,
respectively. The RV curves of the primary and secondary stars are
also over-plotted in the same figure with full and  dotted lines,
respectively. The sinusoidal fit of the RV variation of the B
emission feature is displayed by a dashed line.

\begin{table}
\scriptsize
\begin{center}
  \caption{The EWs and RVs of absorption and emission features in H$\alpha$ difference profiles for S~Equ.}
$$
  \begin{tabular}{cccrr}
  \hline\hline
    HJD         & Orbital   & Type$^{1}$ & EW~      & V$_{r}$~~~       \\
  24 50000+     & Phase     &            & ({\AA})~ & (km s$^{-1}$)  \\
  \hline\hline
  2924.4177 & 0.6182     & Absorption & 0.488   & --61  \\
  2924.4177 & 0.6182     & Emission   & --0.241 & --408  \\
  2924.4177 & 0.6182     & Emission   & --0.160 & 282   \\
  2933.3604 & 0.2207     & Absorption & 0.102   & --163 \\
  2933.3604 & 0.2207     & B          & --0.229 & 20   \\
  2933.3604 & 0.2207     & C          & --0.135 & 125 \\
  3155.5306 & 0.8779     & A          & 0.210   & --43 \\
  3155.5306 & 0.8779     & C          & --0.050 & --180 \\
  3158.5586 & 0.7591     & A          & 0.048   & --30 \\
  3158.5586 & 0.7591     & B          & --0.089 & --121 \\
  3158.5586 & 0.7591     & C          & --0.168 & --244 \\
  3160.5341 & 0.3340     & A          & 0.033   & --98 \\
  3160.5341 & 0.3340     & B          & --0.170 & 16 \\
  3160.5341 & 0.3340     & C          & --0.126 & 121 \\
  3166.5869 & 0.0955     & A          & 0.192   & --73 \\
  3167.5542 & 0.3770     & B          & --0.120 & 1 \\
  3167.5542 & 0.3770     & C          & --0.165 & 74 \\
  3168.5489 & 0.6665     & A          & 0.032   & --23 \\
  3168.5489 & 0.6665     & B          & --0.095 & --110 \\
  3168.5489 & 0.6665     & C          & --0.257 & --225 \\
  3170.5519 & 0.2494     & A          & 0.017   & --87 \\
  3170.5519 & 0.2494     & B          & --0.376 & 36 \\
  3170.5519 & 0.2494     & C          & --0.117 & 137 \\
  3185.5247 & 0.6069     & Absorption & 0.198   & 123  \\
  3185.5247 & 0.6069     & C          & --0.207 & --165  \\
  3186.4374 & 0.8725     & A          & 0.081   & --42 \\
  3186.4374 & 0.8725     & C          & --0.229 & --215 \\
  3202.5972 & 0.5754     & A          & 0.044   & --35 \\
  3202.5972 & 0.5754     & C          & --0.273 & --158 \\
  3203.4302 & 0.8178     & Absorption & 0.037   & 93  \\
  3203.4302 & 0.8178     & B          & --0.108 & --89  \\
  3203.4302 & 0.8178     & C          & --0.194 & --199  \\
  3214.5830 & 0.0636     & A          & 0.147   & --72 \\
\hline\hline
\end{tabular}
$$
\centering
$^{1}$ The features labeled as $"absorption"$ and $"emission"$ have different RV variations from those
listed as A, B and C.
\end{center}
\end{table}

\emph{Extra Absorption Feature (A)}. The visibility and velocity
information, which can be deduced from EW and RV variations,
provide constraints about the location of sources. H$\alpha$ extra
absorption is clearly visible just before and after primary
eclipse (Fig.~4). It can be also easily followed in Fig.~5, which
shows the EWs and RVs of the A-absorption feature. The EWs indicate that
the absorption is absent at phases far from the primary eclipse. The
RVs of this extra-absorption feature closely follow the RV curve of
the hotter component suggesting that it may be connected with the
mass-gainer component (Fig.~6a). This feature could be the result of
circumstellar matter arising from the interaction between the gas
stream and the environment around the hotter star. The EW
variations also indicate that the matter should be denser around the
impact point, where the gas stream interacts with the primary star,
which is more visible around primary eclipse. The strong
extra-absorption observed just before the primary minimum could also be
the effect of the gas stream itself that is projected against the
disk of the hotter component at these phases. However, the extra-absorption
observed just after primary minimum and its RV variation make the impact region
the more likely source of this feature. 
This would also explain the light deficit observed by Catalano \& Rodon\`o (1968)
near 0.9 phase.

\begin{figure}
\begin{center}
\includegraphics[width=80mm,height=45mm]{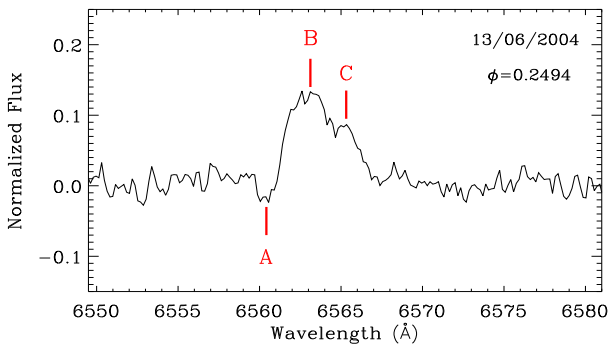}
\caption{A difference H$\alpha$ profile for S~Equ which displays all the
prominent features; the extra-absorption (A) the broad emission component (B)
and the narrower emission connected to the secondary star (C).}
\end{center}
\label{fig:ABC_features}
\end{figure}

\emph{Extra Emission Feature (B)}. The EWs show that the emission
profile (B) is not seen during the eclipses, while it is more
prominent around the quadratures (see Figs.~4 and 6b). Therefore, it
may be related to the circumstellar matter located in the space
between the primary and secondary star, that is occulted at both
eclipses (see Fig.~7). As shown in Fig.~6a, the RVs of the B
emission feature follow the orbital motion of the cooler component
without any significant phase shift but with a smaller amplitude. In
the hypothesis that the emitting region is at rest in the rotating
reference frame of the system, we can deduce that it lies between
the center-of-mass of the system (CM) and the secondary star and we
can also evaluate its location, analogously to what done for HR~7428
by Marino et al. (2001). The projected distance of the region from
the CM can be found as $a_{B}=a_{c}(K_{B}/K_{c})$, where
$a_{c}=a/(1+q)$ is the semi-major axis of the cool star orbit and
$q$ and $a$ are the mass ratio and system separation, respectively.
 Here, $K_{c}$ and $K_{B}$ are the semi-amplitudes of the RV variations
for the cool star and for the emitting region, respectively.
Using the orbital solution of S~Equ given in Table 5
and the semi-amplitude of the RV variation of the B emission
feature estimated as $K_{{B}}\simeq71$\,km\,s$^{-1}$, we find that
the emitting region should be located at about 3$\times10^{6}$ km from the
barycenter toward the cool component. Thus, feature B could be 
part of the gas stream (see Fig.~7).

\begin{figure}
\begin{center}
\includegraphics[width=60mm,height=80mm]{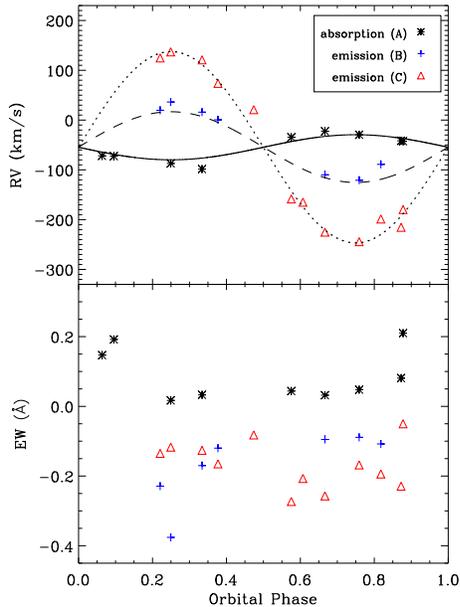}
\caption{The variations of RVs (upper panel) and EWs (lower panel) of extra-absorption
and extra-emission features observed in H$\alpha$ for S~Equ. The continuous and dotted
lines are the RV solutions for the primary and secondary component, respectively.
The dashed sinusoid represents the RV variation of the broad emission region (B)
with a semi-amplitude $K_{B}\simeq71$\,km\,s$^{-1}$. }
\end{center}
\label{fig:ABC_curves}
\end{figure}

\emph{Extra Emission Feature (C)}. Although the emission profile (C)
is visible at nearly the same orbital phases where the B-type
emission is observed, its narrower width (Fig.~5) and its RV
variation (Fig.~6a) clearly indicate that its origin is different.
The velocity of the C feature follows the RV curve of the secondary
companion. Therefore, the emission seems to be closely related to
the cool component and it could be produced by its chromosphere.

A schematic representation of the S~Equ system with the Roche-lobe
geometry, the gas stream, and the extra-emission/absorption regions
is shown in Fig.~7.

\begin{figure}
\begin{center}
\includegraphics[width=75mm,height=78mm]{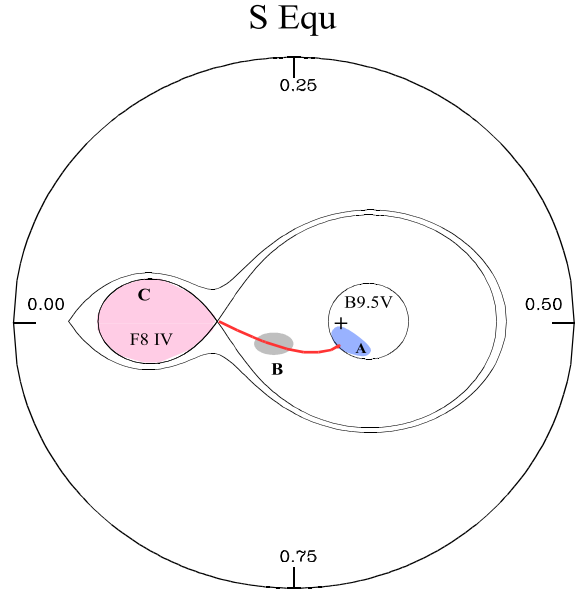}
\vspace{0.5cm}
\caption{Schematic representation of the S~Equ
system with the Roche-lobe geometry, the gas stream, and the
extra-absorption/emission regions (A, B, and C). The barycenter,
which is inside the hotter component, is marked by a cross.}
\end{center}
\label{fig:Roche}
\end{figure}

\subsection[]{KO~Aql}

We have subtracted the synthetic spectrum from each KO~Aql spectrum
using the method described in the previous section in order to
produce residual H$\alpha$ profiles to be analyzed. A sample of
H$\alpha$ profiles is shown in Fig.~8. In the residual profiles is
apparent an absorption feature (A), which is stronger around the
primary eclipse, and an emission profile (B) clearly seen around the
two quadratures. We have measured EWs and RVs of these features to
get information about their origin, as we did for S~Equ. These
values are listed in Table 8 and are plotted against the orbital
phase in Fig.~9. However, features with RV not consistent with A and
B are visible in a few spectra. The EWs and RVs for these features
are not plotted in Fig.~9, but their values are reported in Table~8
and simply labeled as $``absorption"$ or $``emission"$ features.

\begin{figure}
\begin{center}
\includegraphics[width=85mm,height=140mm]{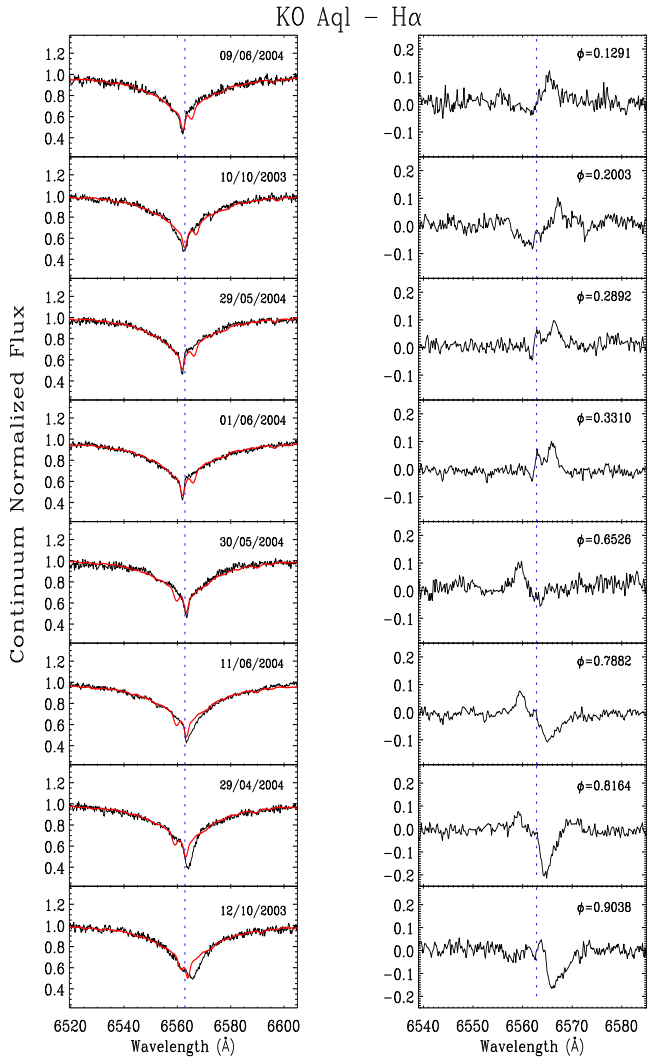}
\caption{\emph{Left.} Sample of observed (thin line) and the synthetic
spectra (thick line) for KO~Aql at different orbital phases. The date of
observation is marked in each panel.
\emph{Right.} The difference spectra showing H$\alpha$ extra-absorption and emission features.
The orbital phase is reported in the upper right corner of each panel.
}
\end{center}
\label{fig:Halpha_KOAql}
\end{figure}

Although the RV of the extra-absorption (A) seems to follow the RV
curve of the mass-gainer (hotter) component at some phases (such as
between 0.25 and 0.35), it displays a much larger amplitude of RV
variation compared to the primary star, especially just before and
after primary eclipse (see Fig.~9). The EWs indicate that the absorption
feature becomes stronger from phase 0.7 to 0.9 and is visible until
phase 0.2. The maximum value of equivalent width,
EW\,$\approx$0.6\,{\AA}, is observed just before primary
eclipse, at about 0.9 phase. This suggests that the absorption is
likely produced by circumstellar matter around the primary component
or to the gas stream itself that is projected onto the hotter star
and would give rise to a red-shifted absorption at these phases, as
it is observed. However, this cannot explain the
blue-shifted absorption observed at about 0.1 phase. Anyway, the
variability and the visibility of the absorption feature (A)
indicate that the source of this extra-absorption may be connected
with the stream-star impact region.

\begin{table}
\scriptsize
 \begin{center}
  \caption{The EWs and RVs of absorption and emission features in H$\alpha$ difference profiles for KO~Aql.}
$$
   \begin{tabular}{cccrr}
 \hline\hline
      HJD     & Orbital   & Type$^{1}$ & EW~        & V$_{r}$~~~       \\
  24 50000+ & Phase     &            & ({\AA})~ & (km s$^{-1}$)  \\
  \hline
2896.3330   & 0.7851    & A       & 0.253      & 59 \\
2896.3330   & 0.7851    & B       & --0.181    & --173 \\
2922.2714   & 0.8416    & A       & 0.438      & 154 \\
2922.2714   & 0.8416    & B       & --0.134    & --146 \\
2923.2988   & 0.2003    & A       & 0.284      & --103 \\
2923.2988   & 0.2003    & B       & --0.148    & 184 \\
2925.3135   & 0.9038    & A       & 0.610      & 164 \\
2925.3135   & 0.9038    & Emission& --0.035    & 11 \\
2933.3237   & 0.7006    & A       & 0.148      & 155 \\
2933.3237   & 0.7006    & B       & --0.143    & --158 \\
2933.3237   & 0.7006    & Absorption & 0.035   & --29 \\
3125.5480   & 0.8164    & A       & 0.551      & 120 \\
3125.5480   & 0.8164    & B       & --0.104    & --152 \\
3126.4985   & 0.1483    & A       & 0.316      & --62 \\
3126.4985   & 0.1483    & B       & --0.145    & 158 \\
3126.4985   & 0.1483    & Emission& --0.239    & 5 \\
3126.4985   & 0.1483    & Emission& --0.108    & 353 \\
3139.5625   & 0.7096    & A       & 0.283      & 91 \\
3139.5625   & 0.7096    & B       & --0.148    & --155 \\
3140.5669   & 0.0603    & A       & 0.242      & --78 \\
3143.4979   & 0.0837    & A       & 0.229      & --127 \\
3143.5831   & 0.1134    & A       & 0.239      & --146 \\
3143.5831   & 0.1134    & B       & --0.149    & 204 \\
3152.4905   & 0.2235    & A       & 0.110      & --49 \\
3152.4905   & 0.2235    & B       & --0.138    & 175 \\
3152.5880   & 0.2575    & A       & 0.068      & --41 \\
3152.5880   & 0.2575    & B       & --0.207    & 179 \\
3155.5687   & 0.2892    & A       & 0.032      & --37 \\
3155.5687   & 0.2892    & B       & --0.216    & 182 \\
3155.5687   & 0.2892    & Emission& --0.070    & 25 \\
3156.5837   & 0.6526    & A       & 0.098      & 57 \\
3156.5837   & 0.6526    & B       & --0.243    & --151 \\
3158.5267   & 0.3310    & A       & 0.034      & --36 \\
3158.5267   & 0.3310    & B       & --0.164    & 161 \\
3158.5267   & 0.3310    & Emission& --0.063    & 29 \\
3166.5406   & 0.1291    & A       & 0.289      & --109 \\
3166.5406   & 0.1291    & B       & --0.256    & 145 \\
3166.5406   & 0.1291    & Absorption & 0.096   & --88 \\
3168.4281   & 0.7882    & A       & 0.223      & 125 \\
3168.4281   & 0.7882    & B       & --0.304    & --152 \\
3168.5833   & 0.8423    & A       & 0.433      & 144 \\
3168.5833   & 0.8423    & B       & --0.137    & --142 \\
3171.5641   & 0.8831    & A       & 0.323      & 151 \\
3171.5641   & 0.8831    & B       & --0.120    & --111 \\
3216.5501   & 0.5901    & A       & 0.144      & 170 \\
3216.5501   & 0.5901    & Absorption & 0.063   & --189 \\
\hline\hline
\end{tabular}
$$
\centering
$^{1}$ The features labeled as $"absorption"$ and $"emission"$ have different RV variations from those
listed as A, B and C.
\end{center}
\end{table}

\begin{figure}
\begin{center}
\includegraphics[width=60mm,height=80mm]{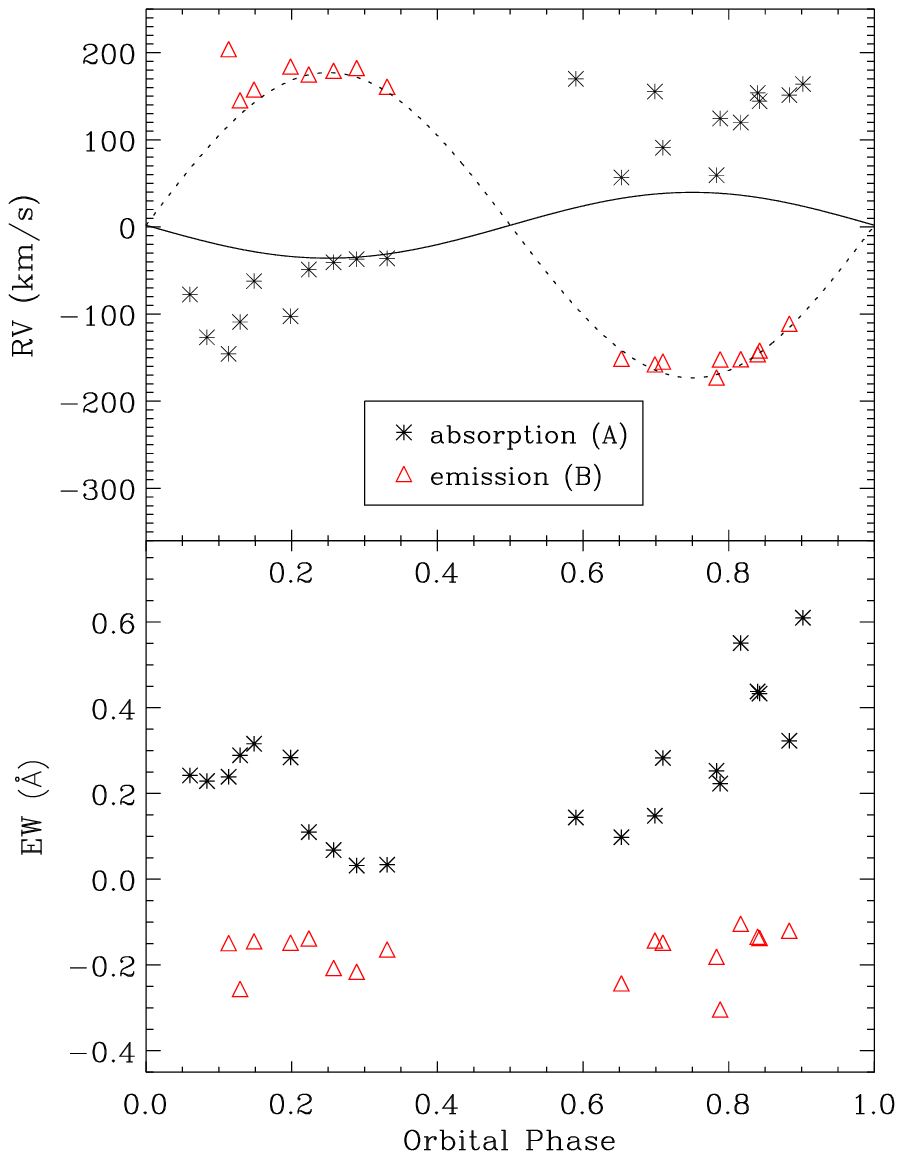}
\caption{The variations of RVs and EWs of H$\alpha$ difference profiles for KO~Aql.
As before, the continuous and dotted lines represent the RV solutions for the
primary and secondary, respectively.}
\end{center}
\label{fig:ABC_KOAql}
\end{figure}

To estimate the location and the origin of the extra-emission
feature (B), we examine the variation of its RVs and EWs in Fig.~9.
The EWs have a value of about 0.2~{\AA} and no relevant change with
orbital phase. The RV of this emission feature closely follows
that of the secondary star, indicating a possible chromospheric
origin. Unlike S~Equ, for KO~Aql there is no clear evidence
of a broad extra-emission feature attributable to the mass transfer
phenomenon. A schematic representation of the KO~Aql system with the
Roche-lobe geometry, the gas stream, and the
extra-emission/absorption regions is shown in Fig.~10.

\begin{figure}
\begin{center}
\includegraphics[width=75mm,height=78mm]{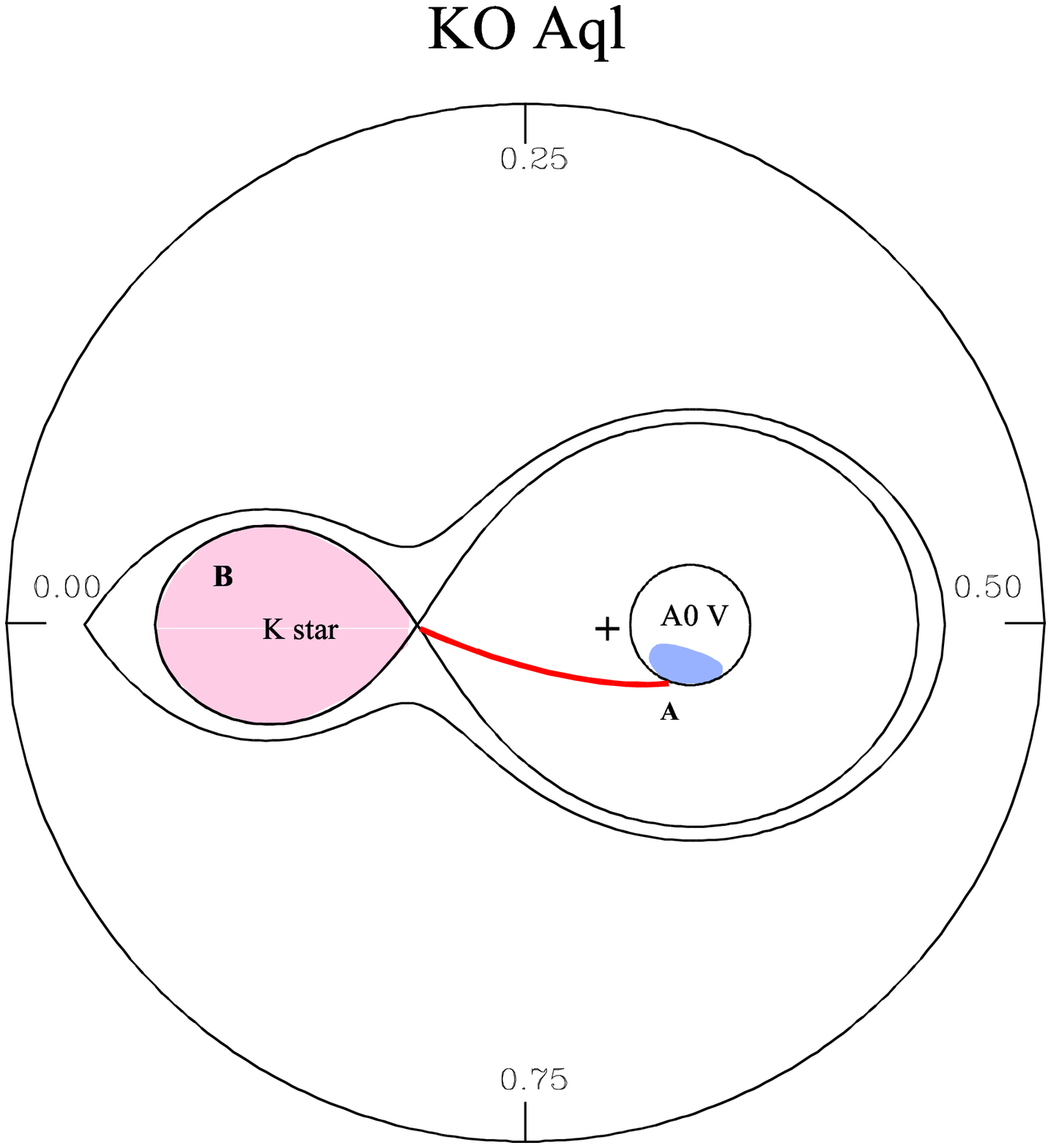}
\caption{Schematic representation of the KO~Aql system with the Roche-lobe geometry,
the gas stream, and the extra-absorption/emission regions (A and B). The barycenter,
close to the hotter component, is marked by a cross.}
\end{center}
\end{figure}

\section[]{Discussion}

According to the catalogue compiled by Budding et al. (2004), the number of Algol-type
binaries with RV curves for both components and solved orbits is very limited.
Accurate values of stellar and orbital parameters for as many binaries as possible
are strongly needed for the study of several phenomena such as mass transfer and
accretion, magnetic activity on late-type components, and angular
momentum evolution.

In this study, we have presented new radial velocity curves for the hotter primary
components of S~Equ and KO~Aql with much better precision and phase coverage
than previous works. Moreover, we have obtained the first RV curves for the 
late-type secondary components of both systems.

Thus, the first complete orbital solutions and mass determination have been made
for both systems. From the photometric and spectroscopic elements given in Tables 1
and 5, respectively, we derived the absolute parameters of S~Equ and KO~Aql, which
are listed in Table 9. The masses are determined to better than 2$\%$, and the radii
with a 4$\%$ precision.

From the effective temperatures and radii we have calculated the
luminosities of the components of both systems and their absolute
bolometric magnitudes, adopting for the
Sun $M_{\rm bol}^{\sun}$\,=4.74 \citep{Cox00}.  
The absolute magnitudes in the $V$ band, $M_V$, were calculated from
the bolometric magnitudes by applying the bolometric corrections
appropriate for the temperatures of our stars (Flower 1996). The
{\it combined} absolute magnitude (including both components) is
$M_{V}^{\rm tot}=0\fm095$ for S~Equ and $M_{V}^{\rm tot}=1\fm326$
for KO~Aql. We have also calculated the {\it combined} $B-V$ color
expected from the sources by means of the calibration $T_{\rm eff}$
versus $B-V$ provided by Flower (1996), finding
$(B-V)_0^{tot}=-0\fm032$ for S~Equ and $(B-V)_0^{tot}=+0\fm046$ for
KO~Aql.

From the absolute and apparent $V$ magnitudes we can easily derive the distance modulus,
if we know the reddening to the sources.

For S~Equ, we have used the outside-of-eclipse values $V=8\fm37$,
$B-V=0\fm072$ (ESA 1997), that lead to a color excess $E(B-V)\simeq
0\fm104$ and an extinction $A_V\simeq 0\fm324$, with
$R=A_V/E(B-V)=3.1$ \citep{SavMat79}. A distance of 390\,$\pm$\,40 pc
is found, in excellent agreement with the astrometric determination
of 400\,$\pm$\,210 pc from the \emph{Hipparcos} mission (ESA, 1997).

For KO~Aql, with $V=8\fm42$ and $B-V=0\fm114$ (ESA 1997), we find $E(B-V)\simeq 0\fm068$ and
$A_V\simeq 0\fm212$. The distance derived in the same way is 240\,$\pm$\,20 pc, in  good agreement
with the astrometric distance of 265\,$\pm$\,75 pc from the \emph{Hipparcos} mission (ESA, 1997).

\begin{table*}
\caption[]{Absolute parameters and related quantities for S~Equ
and KO~Aql. $v\sin i$ is the observed projected rotational velocity for both components.
$v_{\rm syn}\sin i$ is instead the projected rotational velocity of the primary components calculated
assuming synchronous rotation.}
 \label{Table2}
$$
\begin{tabular}{lcccccccc}
\hline\hline
                   & & \multicolumn{3}{c}{S~Equ} & & \multicolumn{3}{c}{KO~Aql} \\
Parameters         & & Primary & &  Secondary  & & Primary & &  Secondary \\
\cline{1-1}\cline{3-5}\cline{7-9}\\
Mass (M${_{\sun}}$)    & & 3.24 $\pm$ 0.03 & & 0.42 $\pm$ 0.01 & & 2.53 $\pm$ 0.05 & & 0.55 $\pm$ 0.01  \\
Radius (R${_{\sun}}$)  & & 2.74 $\pm$ 0.09 & & 3.24 $\pm$ 0.10 & & 1.74 $\pm$ 0.07 & & 3.34 $\pm$ 0.07  \\
log T (K)              & & 4.049$^{a}$     & & 3.72$^{a}$      & & 3.996$^{b}$     & & 3.646$^{b}$      \\
log L (L${_{\sun}}$)   & & 2.02 $\pm$ 0.10 & & 0.86 $\pm$ 0.10 & & 1.41 $\pm$ 0.06 & & 0.56 $\pm$ 0.06  \\
log g (cgs)            & & 4.07 $\pm$ 0.01 & & 3.04 $\pm$ 0.01 & & 4.36 $\pm$ 0.03 & & 3.13 $\pm$ 0.01  \\
a (R${_{\sun}}$)       & & \multicolumn{3}{c} {14.8 $\pm$ 0.1} & & \multicolumn{3}{c} {12.05 $\pm$ 0.05}\\
$v_{\rm syn}\sin i$ (km/s) & & 40.3            & & 47.7            & & 30.1            & & 57.7             \\
$v\sin i$ (km/s) & & 52.0 $\pm$ 4.0  & & ...             & & 41.0 $\pm$ 2.0  & & ...              \\
M$_{bol}$ (mag)        & & $-0.32 \pm$ 0.11 & & 2.60 $\pm$ 0.15 & & 1.20 $\pm$ 0.13 & & 3.28 $\pm$ 0.15  \\
M$_{v}$ (mag)          & & 0.19 $\pm$ 0.11 & & 2.81 $\pm$ 0.15 & & 1.43 $\pm$ 0.13 & & 3.94 $\pm$ 0.15  \\
Distance (pc)          & & \multicolumn{3}{c} {390 $\pm$ 40} & & \multicolumn{3}{c} {240 $\pm$ 20}\\
\hline\hline
\end{tabular}
$$
\centering
$^{a}$Zola (1992), $^{b}$Mader \& Angione (1996).
\end{table*}

The positions of the primary and secondary components of S~Equ and
KO~Aql on the Hertzsprung-Russel (HR) diagram are shown in Fig.~11,
together with the theoretical evolutionary tracks for the more
massive components, as computed by Girardi et al. (2000), and with
the conservative evolutionary track for the mass-gainer star
belonging to a system with a total mass of $M_1+M_2 = 4.2\,M_{\sun}$
and an initial orbital period of
1.2 days calculated by De Loore \& Van Rensbergen (2005)\footnote{\texttt http://we.vub.ac.be/astrofys/binarytracks/index.html}.    

\.{I}bano\v{g}lu et al. (2006) have pointed out that semi-detached
systems (SDBs) display relevant differences compared to the detached
binaries (DBs). The primary components, mass-gainers of SDBs, seem
to be normal main-sequence stars. The cooler ones, the losers, are
instead stars evolved off the main sequence which are all
oversized and over-luminous for their masses. A relevant difference
between the primary components of SDBs and DBs has been found by
\.{I}bano\v{g}lu et al. (2006) who determined the mass-luminosity
relation (MLR). They have taken into account 74 DBs and 61 SDBs with
well-determined absolute parameters and derived for the primaries of
SDBs a MLR relation L$_{1} \propto M_{1}^{3.20}$, while the
luminosities of the primaries of DBs are correlated with their
masses as M$_{1}^{3.92}$. These relations show that the hotter
components of SDBs are under-luminous relative to the main-sequence
stars of the same-mass.

This trend is also followed by the hotter components of S~Equ and KO~Aql which are
clearly underluminous compared to normal main-sequence stars of the same mass (Fig.~11).
This behaviour is not dependent on the theoretical tracks, because we find exactly the same results
by adopting the evolutionary tracks of Pols et al. (1998), and so the trend could be due to the mass
that is being acquired by these stars by Roche-lobe overflow from the donor.

Indeed, the evolutionary track of the mass-gaining component for an Algol system with
$M_1+M_2 = 4.2\,M_{\sun}$ (dotted line in Fig.~11) shows that, if the mass-transfer is
still ongoing, the position of the hotter component can be significantly different
from that of a normal main-sequence star with the same mass. Unfortunately, De Loore
\& Van Rensbergen (2005) did not calculate evolutionary tracks for systems with
the same total mass as S~Equ (3.66$\,M_{\sun}$) and KO~Aql (3.03$\,M_{\sun}$), but this
result indicates that the luminosity discrepancy could be explained by an evolutionary effect.

\begin{figure}
\begin{center}
\includegraphics[width=85mm]{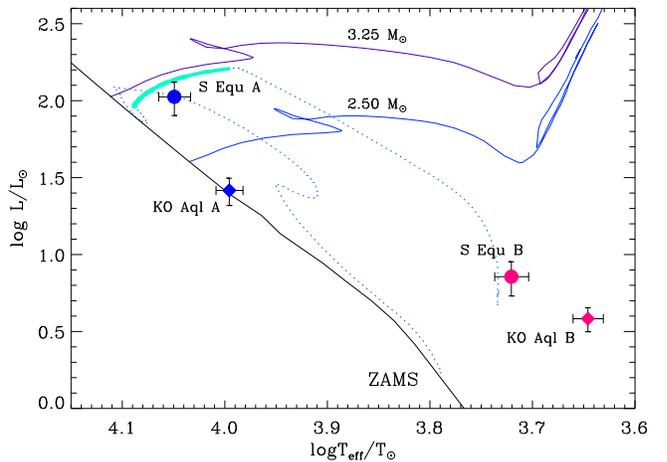}
\caption{HR diagram of the primary (A) and secondary (B) components of S~Equ (dots) and KO~Aql (diamonds).
The ZAMS and  the post-main-sequence evolutionary tracks (full lines) for the masses of the
primary components of S~Equ and KO~Aql from Girardi et al. (2000) are also shown for comparison.
The evolutionary track of the mass-gainer component of a semi-detached system with a total
mass of 4.2\,$M_{\sun}$ according to De Loore \& Van Rensbergen (2005) is displayed with a dotted line. The thick grey line
marks the portion of this track for which the mass of the gainer is comprised between 3.1 and 3.4\,$M_{\sun}$.
}
\end{center}
\label{fig:HR}
\end{figure}

S~Equ and KO~Aql are members of the group of short-period Algols
(P$_{orb}$ < 5--6 days), in which the mass transfer and accretion
process are more complicated than in the long-period Algols
(P$_{orb}$ > 5--6 days) because the gas stream from the donor to the
gainer directly strikes the star. This stream-star interaction
results in a variety of accretion features. Generally, an accretion
structure in these systems, which is called an \emph{accretion
annulus} (Richards et al. 1996), can be produced. This structure is
an asymmetric, sometimes clumpy distribution of gas around the
primary component (Richards 2001). The observed properties of the
mass transfer and circumstellar material in short-period Algols have
been mostly found from spectroscopic studies, especially from
variations of Balmer-line profiles. The identified regions include
the gas stream (Richards et al. 1995), a transient accretion disk
(Kaitchuck \& Honeycutt 1982), and a denser region near the impact
site called a \emph{localized region} (Richards 1992, Richards et al.
2000).

\begin{figure}
\begin{center}
\includegraphics[width=75mm,height=60mm]{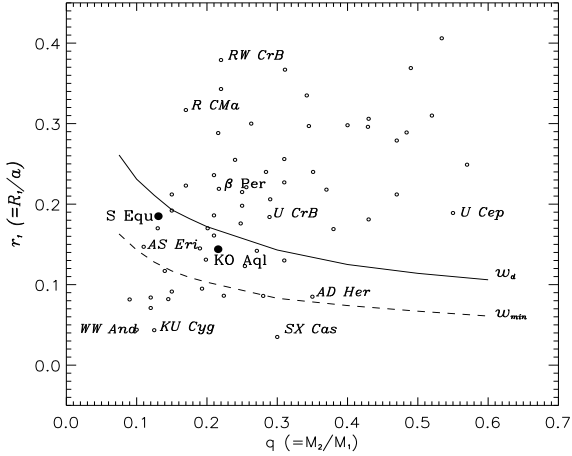}
\caption{The \emph{r-q} diagram for classical Algols, where \emph{r}
is the fractional radius of the primary star and \emph{q} is the
mass ratio of the system. The curve of $\omega_{min}$ represents the
minimum distance between the gas stream and the center of the
primary star, while the curve of $\omega_{d}$ corresponds to the
distance of closest approach of the stream. Notice that S~Equ and KO
Aql lie between the two curves, indicating unstable and transient
accretion structures. }
\end{center}
\label{fig:omega}
\end{figure}

The type of accretion structures can be predicted by using the
ballistic calculations of Lubow \& Shu (1975) and the $r_1-q$
diagram, a plot of radius of the mass gainer versus the mass ratio
($q=M_{2}/M_{1}$) of the system (Peters 1989). Two curves
($\omega_{d}$ and $\omega_{min}$) can be drawn onto the diagram,
where $\omega_{d}$ defines the fractional radius of the outer disk
and $\omega_{min}$ represents the minimum distance between the
stream and the center of the primary star.

We have constructed the $r_1-q$ diagram (Fig.~12) for all the SDBs
from the list of \.{I}bano\v{g}lu et al. (2006), taking $r_1$ and
$q$ from the catalog of Budding et al. (2004). The positions of the
gainers of S~Equ and KO~Aql are marked with big dots. The
short-period systems (P$_{orb}$ < 5--6 days) are located above the
$\omega_{d}$ curve; the gas stream strikes the primary directly,
resulting into an impact region and possibly a transient disk or
annulus because the orbital separation of two components is too
small to form a stable disk. Conversely, the systems falling under
the $\omega_{min}$ curve are long-period ones with a small
fractional radius $r_1$ and can form a classical accretion disk
because the stream does not hit the primary. For the systems lying
between the two curves, the accretion structure around the primary
is very unstable and transient (Peters 1989, Richards \& Albright
1999). S~Equ and KO~Aql belong to this group, thus they should
contain transient, unstable disks or an accretions annulus around the
gainers and show evidence of impact regions.

Indeed, our H$\alpha$ difference profiles suggest accretion
structures more closely connected to the gas stream or to the impact
region of the stream on the outer atmosphere of the hot star rather
than to disks. The extra-absorption observed in the S~Equ spectra
follows the motion of the hotter component and is close to the
star-stream impact site, as already found by Richards (2004) who
also found evidence of a circum-primary equatorial `bulge' centered
on the velocity of the primary star. The emission component (B)
found by us displays instead a different behaviour, being likely
located between the two stars and possibly connected to the gas
stream itself or to the interaction between the stream and a
transient disk. This seems to indicate a variability of the
mass-transfer process in this system, in agreement with the
observation of Algol systems that alternate between stream-like and
disk-like states (Richards 2004). The chromospheric emission of the
cool component of S~Equ is also detected in our spectra, as already
found by Richards (2001). We have detected similar features in the
residual H$\alpha$ spectra of KO~Aql, with the exception of the
excess B emission between the two components. In addition, the
extra-absorption displays larger velocity variations compared to
S~Equ.

In Algol systems, the rotational velocity of the mass-gaining
components is an indicator of mass transfer, which acts as a spin-up
mechanism (Olson \& Etzel 1994, Mukherjee et al. 1996). Thus, when
one sees a primary component rotating faster than expected from 
tidally-induced synchronism, one can argue that mass transfer has
been taking place. However, this is not a direct proof for an
ongoing mass flow. By the cross-correlation of the spectra of S~Equ
and KO~Aql with slowly-rotating templates, we have measured, with a
rather good accuracy (2--4 km\,s$^{-1}$), the rotational velocities
of the hotter components of both systems and found that these
components are rotating about 30\% faster than the synchronized
velocity which is likely the result of the accretion of gas with
high angular momentum.

The mass transfer rates in Algols can be estimated assuming a
conservative mass and angular momentum transfer, i.e. without losses
from the system. In such hypothesis, a monotonic increase of the
orbital period is expected. An increasing $P_{\rm orb}$ has been
indeed found for S~Equ (Soydugan et al. 2003) and KO~Aql (Soydugan
\& Soydugan 2007). The analysis of the orbital period changes has
allowed us to calculate the mass transfer rates $dM/dt\simeq
4.0\times 10^{-8}M_{\sun} yr^{-1}$ for S~Equ and $dM/dt\simeq
2.6\times 10^{-7}M_{\sun} yr^{-1}$ for KO~Aql. Considering that
Algols are mostly in the slow phase of mass transfer with rates
$dM/dt\sim 10^{-11} - 10^{-7} M_{\sun} yr^{-1}$ (Richards $\&$
Albright 1999), the mass transfer rates for S~Equ and KO~Aql are
among the highest occurring in these systems, thus explaining the
strong H$\alpha$ extra-absorption and/or emission features shown by
these systems.

\section[]{Conclusions}

We have observed spectroscopically the classical Algol-type binaries S~Equ and KO~Aql
with the aim of determining the physical parameters of the components and to study
the mass transfer and accretion structures around the hotter components by
analyzing H$\alpha$ difference profiles.

Our detailed spectroscopic analysis has yielded the first precise
determinations of the absolute parameters of the systems and, in
particular, the stellar masses, thanks to the first radial velocity
curves of the cooler secondary components. We found that the hotter,
mass-gaining components of both systems are underluminous compared
to normal main-sequence stars of the same mass. We show that this
behaviour can be explained by the Roche-lobe overflow.

We have also found clear evidence for mass transfer, accretion
structures, and chromospheric activity of late-type components in
our spectra. In particular, we have detected in S~Equ a broad
H$\alpha$ extra-emission component that is observed red-shifted at
phases around 0.25 and blue-shifted at phases near 0.75. This
feature is not observed during the eclipses and its RV variation is
consistent with an accretion structure (mass flow or interaction
between the stream and a transient disk) located between the two
stars at about 3$\times10^{6}$\,km from the barycenter toward the
cool component. Similar features were also observed by Vesper et al.
(2001) in some short-period Algol binaries such as SS~Cet, U~CrB,
$\beta$ Per, and S~Equ itself. However, we did not observe a similar
behaviour in KO~Aql.

Our H$\alpha$ spectra also display, for both S~Equ and KO~Aql, extra-absorption features which get stronger
around phase 0 and whose radial velocities are consistent with those of the hotter components.
They are very likely produced by the impact region between the stream and the primary.
Analogous features were already observed in other short-period Algols such as $\beta$~Per
(Richards 1993), SW~Cyg (Hunt 1997) and RW~Mon (Vesper \& Honeycutt 1999).

In the difference profiles of both systems we have also detected emission lines which closely
follow the radial velocity curves of the late type secondaries and, as such, they very likely
arise from their chromospheres.

\textsf{Acknowledgements}. {We are grateful to the referee, Dr. Mercedes Richards, for several 
helpful comments and a careful reading of the manuscript. F. Soydugan would like to thank all the
staff of Catania Astrophysical Observatory for the kind hospitality,
the allocation of telescope time, and the precious help during the observations.
He is also grateful to the Scientific and Technical Research Council
of Turkey (T\"{U}B\.{I}TAK) for their financial support.
We are grateful to Dr. Walter Van Rensbergen for very useful suggestions.
This research has been partially supported by INAF ({\em Istituto Nazionale di Astrofisica}) and
Italian MIUR ({\em Ministero dell'Istruzione, Universit\`a e Ricerca}).
This research has made use of SIMBAD database.}

\bsp

\label{lastpage}

\end{document}